\documentclass[prb,
twocolumn,
superscriptaddress,showpacs,amsmath,amssymb,longbibliography]{revtex4-2}

\usepackage{xcolor}
\usepackage{graphicx}
\usepackage{hyperref}

\definecolor{TTH-color}{rgb}{0.0,0.0,1}
\definecolor{GV-color}{rgb}{1,0,0}
\definecolor{JN-color}{rgb}{255,0,255}

\begin{document}

\title{Topological polarization, dual invariants, and surface flat band in crystalline insulators}

\author{J. Nissinen}
\affiliation{Low Temperature Laboratory, Aalto University,  P.O. Box 15100, FI-00076 Aalto, Finland}

\author{T. T. Heikkil\"a}
\affiliation{Department of Physics and Nanoscience Center, University of Jyv\"askyl\"a, P.O. Box 35 (YFL), FI-40014 University of 
Jyv\"askyl\"a, Finland}

\author{G.E.~Volovik}
\affiliation{Low Temperature Laboratory, Aalto University,  P.O. Box 15100, FI-00076 Aalto, Finland}
\affiliation{Landau Institute for Theoretical Physics, acad. Semyonov av., 1a, 142432,
Chernogolovka, Russia}

\date{\today}

\begin{abstract}
{
We describe a three-dimensional crystalline topological insulator (TI) phase of matter that exhibits spontaneous polarization. This polarization results from the presence of (approximately) flat bands on the surface of such TIs. These flat bands are a consequence of the bulk-boundary correspondence of polarized topological media, and contrary to related nodal line semimetal phases also containing surface flat bands, they span the entire surface Brillouin zone. We also present an example Hamiltonian exhibiting a Lifshitz transition from the nodal line phase to the TI phase with polarization. Utilizing elasticity tetrads, we show a complete classification of 3D crystalline TI phases and invariants. The phase with polarization naturally arises from this classification as a dual to the previously better-known 3D TI phase exhibiting quantum (spin) Hall effect. Besides polarization, another implication of the large surface flat band is the susceptibility to interaction effects such as superconductivity: the mean-field critical temperature is proportional to the size of the flat bands, and this type of systems may hence exhibit superconductivity with a very high critical temperature.
}
\end{abstract}
\pacs{
}

\maketitle 

\section{Introduction}

The best-known topological insulators \cite{Hasan2010} in two dimensions are characterized by robust edge states and a (spin) Hall conductivity \cite{qhenote1} quantized in the units of $\sigma_0=e^2/h$. In three dimensions, conductivity scales like  $\sigma_0/[\ell]$, where $[\ell]$ is a length scale characteristic to the system under consideration. In crystalline matter, the relevant length scale is obtained from the lattice vectors \cite{Halperin1987,roy2009topological} and the quantum Hall response is different in different directions specified by the reciprocal lattice vectors.

Another type of a topological response is the electric polarization. It has been discussed e.g. in Refs.~\onlinecite{KingSmith1993, Resta1994,Resta2010,Nomura2018, Armitage2019}, and it has recently attracted renewed interest{\cite{Rhim2017, Miert2017, Sergeev2018, Murakami2020, BudichArdonne2013}}. The quantized quantity is the 2D polarization charge density that scales like $1/[\ell^2]$. In crystalline media we may hence expect the relevant length scales to be associated with the crystal lattice vectors. 

A natural framework to describe the topological response in crystalline media is in terms of elasticity tetrads $E^{\ a}_{\mu} = \partial_{\mu} X^a$, where $X^a$ counts the number of lattice planes along crystal direction $a$. They are a convenient way to discuss {semi-classical hydrodynamics, elastic deformations and conserved charges in terms of {(continuum)} lattice geometry} \cite{DzyalVol1980}. Different to the dimensionless tetrads in the first order formulation of general relativity, they have the canonical dimensions of inverse length {inherited from the underlying lattice}. {More recently, they have been shown to enter the {field theoretical} topological response of crystalline insulators with additional conserved lattice charges, specified by integer quantized momentum space invariants \cite{NissinenVolovik2019,Vishwanath2019}}, and they can be extended \cite{Volovik2020} to the relativistic quantum fields and gravity \cite{Wetterich2005, Diakonov2011,VladimirovDiakonov2012,ObukhovHehl2012, Wetterich2012, VladimirovDiakonov2014}. {In these cases, the associated crystal lattice is not necessarily due to periodic real space structure on which the fermionic system is placed but can be induced by interactions and/or other superstructures in the relevant ground state \cite{Oshikawa2000}.}

\begin{figure}[h]
\centering
\includegraphics[width=0.81\columnwidth]{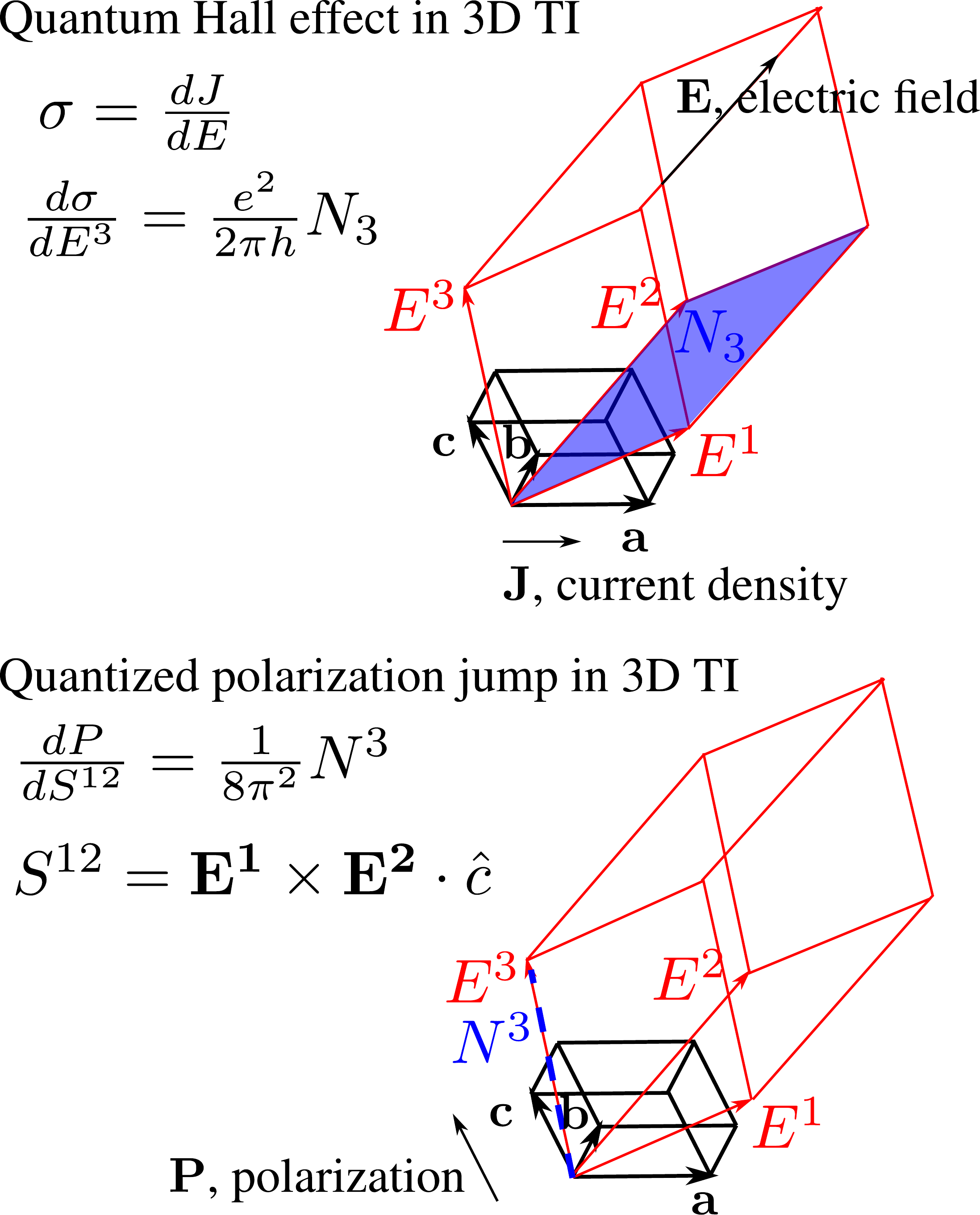}
\caption{Elementary cells of a crystal in real (spanned by ${\mathbf a}$, $\mathbf b$, $\mathbf c$) and reciprocal spaces ($E^a$). Quantum Hall effect \cite{qhenote1,qhenote2} is determined by the topological charge $N_a$ integrated over the surface spanned by a pair of $E^a$, whereas the polarization jump is described by the charge $N^a$ integrated along one $E^a$.}
\label{fig:invariants}
\end{figure}

Here we show that using a combination of the elasticity tetrads $E_\mu^a$ and the electromagnetic gauge fields $A_\mu$, one can in 3+1 dimensions construct four different types of topological terms in the electromagnetic action. They are presented in Eqs.~\eqref{eq:volume_charge}, \eqref{eq:theta_term}, \eqref{action} and \eqref{actionPolarization}, and correspond to dual responses containing different number of elasticity tetrads: three and zero or one and two elasticity tetrads, respectively. As we discuss in Sec.~\ref{subs:0D3D}, the first two describe the trivial band insulator and axion electrodynamics, respectively. The third term describes the 3D quantum Hall effect \cite{NissinenVolovik2019}, and the fourth the topological polarization. The latter two phases are schematically described in Fig.~\ref{fig:invariants}. 

Besides quantized bulk polarization, the TI phase with polarization implies protected boundary modes. There is a marked difference between the boundary modes of the QHE-type topological insulators and those exhibiting surface polarization. Namely, whereas the previous form chiral "half-Dirac" surface modes \cite{Fu2007}, the surface states in the latter case form approximate flat bands spanning the entire surface Brillouin zone (BZ).  Similar flat bands are found in the surfaces of nodal line semimetals,\cite{Heikkila2011a,Heikkila2011,Kopnin2011,Burkov2011} nodal line superconductors\cite{Schnyder2011} and superfluids \cite{Eltsov2019}. However, in those cases the flat bands span only part of the surface BZ, corresponding to the projection of the nodal lines to the surface.
 
{Accordingly}, here we discuss the topological polarization and flat band using an extension of a simple model\cite{Heikkila2011a} to {the range of parameters relevant for crystalline insulators (or superconductors)}. In this extension the multiple Dirac points in a {layered} quasi-2D system evolve into a flat band, which occupies the whole 2D BZ on the boundaries of the 3D system when the number of atomic layers increases. This is accompanied by the formation of a topological {crystalline} insulator state in the bulk. {In the numerical model, we consider the topological response and the corresponding topological invariants for the bulk topological insulator in terms of the elasticity tetrads and calculate the generalized polarization, matching the polarization by the surface flat bands}.

\section{Topological polarization and dual invariants in crystalline insulators}

In this section, we first review the elasticity tetrads, following Ref. \cite{NissinenVolovik2019}, representing {continuum} translational gauge fields {in the crystalline system}. We then discuss the dual forms of topological responses arising from the elasticity tetrads in crystalline insulators in three dimensions. These are, respectively, the total charge conservation and theta term and the three-dimensional quantum Hall effect and topological polarization. Although charge transport is suppressed by the mobility gap in insulators, stricly speaking, the system is not gapped  since e.g. the elasticity tetrads explicitly include symmetry breaking Goldstone modes. This carries over to the responses that are quantized in terms of the invariants and combinations of elasticity tetrads. Throughout the paper, we work mostly in units where $\hbar = e = 1$.

\subsection{Elasticity tetrads}
Let us consider the theory of {crystalline} elasticity using the approach of Refs.~\onlinecite{DzyalVol1980,NissinenVolovik2019}. {An arbitrary, weakly} deformed crystal structure can be described as a system of three crystallographic surfaces, Bragg planes, of constant phase $X^a(x)=2\pi n^a$, $n^a \in \mathbb{Z}$ with $a=1,2,3$. The intersection of the surfaces
\begin{equation}
X^1({\bf r},t)=2\pi n^1 \,\,, \,\,  X^2({\bf r},t)=2\pi n^2 \,\,, \,\, X^3({\bf r},t)=2\pi n^3 \,,
\label{points}
\end{equation}
{represent the lattice points of a deformed crystal}. {In the continuum limit,} the elasticity tetrads are  gradients of the phase functions:
\begin{equation}
E^{~a}_i(x)= \partial_i X^a(x)\quad i=x,y,z, \quad a=1,2,3,
\label{reciprocal}
\end{equation}
{for a three-dimensional spatial crystal. {For simplicity, we work with the orthorombic unit cell lattice system, but the generalization to other lattice symmetries and bases is straightforward and does not affect the general results}. Generalizing to temporal directions}, in an equilibrium {(spacetime)} crystal lattice the quantities $E^{\ a}_\mu$ are lattice {four}-vectors of the reciprocal {(four-dimensional)} Bravais lattice.  Here $E^a_t$ would describe dynamic changes in the lattice, such as phonons, whereas $E_\mu^0$ would correspond to a periodicity in time. In what follows, we concentrate on the static case, but the formulas are readily generalizable to the dynamic case as well. In a deformed crystal, but in the absence of dislocations the tetrads $E^a_\mu$ satisfy the integrability condition of vanishing torsion \cite{NissinenVolovik2019}
\begin{equation}
T^a_{\mu\nu} = \partial_\nu E^{\ a}_\mu-\partial_\mu E^{\ a}_\nu=0\,.
\label{intergrability}
\end{equation}
{These tetrads have dimension of inverse length, $[E^a_\mu]=1/[l]$, being gradients of dimensionless functions $X^a$}.
{This and the presence of finite lattice symmetries} is the main difference to the dimensionless tetrads used in the theories of general relativity.
However, {also in some theories of gravity,} the tetrads have {naturally} dimension $1/[l]$, see {e.g.} Refs.~\onlinecite{Wetterich2005, Diakonov2011,VladimirovDiakonov2012,ObukhovHehl2012, Wetterich2012, VladimirovDiakonov2014,Volovik2020}. 

Moreover, due to the periodicity of the crystal, the functions $X^a$ play the role of continuum $U(1)$ fields, and thus the tetrads play the role of {tautological} vector potentials {representing} effective gauge fields {corresponding to conserved lattice charges in different directions. In a crystalline topological insulator/superconductor, their products correspond to the approximate low-energy (higher-form) symmetries of charge conservation along lattice lines or surfaces below the mobility/quasiparticle gap \cite{Nissinen2020}. In this way, the higher-dimensional {bulk state} can still be (weakly) topologically non-trivial with associated {non-zero and quantized} momentum space invariants in the response. For similar ideas, see \cite{Tuegel2019, Dubinkin2020}. The remaining elasticity tetrad fields enter in new crystalline topological terms 
and contain a mixture of the electromagnetic  $A_\mu$ and elastic  $E^a_\mu$ gauge fields \cite{Nissinen2020}, {as we next discuss}.

{\subsection{Lattice volume and 3D theta term}
\label{subs:0D3D}
To set the stage, we first discuss the tautological topological conservation law of lattice charges in an insulator and its dual response. The three-dimensional lattice volume form is
\begin{align}
\textrm{volume(BZ)} = \mathbf{E}^{1} \wedge \mathbf{E}^{2} \wedge \mathbf{E}^3 \label{eq:BZvol}
\end{align}
Related to this, the insulator has an integer number {of filled electronic bands per unit cell, (minus the positive background charge) and no free charges below the mobility gap \cite{Oshikawa2000}}. The charge density \eqref{eq:BZvol} couples to the electric potential $A_0$ as \cite{Vishwanath2019, NissinenVolovik2019} via the topological action
\begin{align}
S_{0D}=\frac{1}{(2\pi)^3}\int d^4x N_\omega\epsilon^{\mu\nu\lambda\rho} E^{1}_{\mu} E^{2}_{\nu} E^3_{\lambda} A_{\rho} \label{eq:volume_charge}
\end{align}
where the invariant in terms of the (semiclassical) Green's function $G=G(\mathbf{p},\omega)$ is
\begin{align}
N_\omega(\mathbf{p}) = \frac{1}{2\pi i}\int_{-\infty}^{\infty} d\omega \textrm{Tr} G \partial_\omega G^{-1} 
\label{Nw1}
\end{align}
counts the number of occupied states in the BZ. The invariant $N_\omega \equiv N_\omega(\mathbf{p})$ can only change when the gap in the spectrum closes. This makes $N_\omega$ the simplest topological invariant possible. Note that the lattice vectors represented by the elasticity tetrads carry spatial indices only, therefore singling out the potential $A_0$.  While the conservation of lattice volume is tautological to charge conservation {below the (mobility) gap}, and must be compensated by overall charge neutrality over the unit cell, the response \eqref{eq:volume_charge} can be non-trivial when considered on the surface of a topological state with polarization \cite{Vishwanath2019}, see below. This arises since the boundary response is dictated by overall conservation laws from non-trivial higher-dimensional bulk terms and often is anomalous as a purely lower-dimensional theory. 

The response, thus understood, applies to insulators under elastic deformations, i.e. with coordinate dependence on the tetrads $\mathbf{E}^a$. The dual topological response corresponding to the lattice volume is the bulk theta term, corresponding to axion electrodynamics \cite{Qi2008}, 
\begin{align}
S=\frac{1}{32\pi^2}\int d^4 x N_\theta(x,t) \epsilon^{\mu\nu\lambda\rho} F_{\mu\nu} F_{\lambda \rho} \label{eq:theta_term}
\end{align}
coupling to zero-dimensional lattice points, i.e. the response of the {original point charges}. Here $F_{\mu\nu} = \partial_\mu A_\nu-\partial_\nu A_\mu$ is the electromagnetic tensor and
\begin{align}
N_\theta(x,t) &= \frac{1}{96\pi^2}\int_0^{2\pi} du \int_{\rm BZ} d\omega d^3 \mathbf{p} \epsilon^{u\mu\nu\lambda\rho} \textrm{Tr} \big[(G\partial_u G^{-1}) \\ 
&\times (G\partial_{\mu}G^{-1})(G\partial_{\nu}G^{-1})((G\partial_{\lambda}G^{-1})(G\partial_{\rho}G^{-1})\big] \nonumber
\end{align}
with $\partial_{\mu} = (\partial_{\omega}, \partial_{\mathbf{p}})$ is the invariant corresponding to the whole frequency-momentum space, extended by the periodic adiabatic parameter $u$ \cite{Volovik1988,Qi2008,Vayrynen2011,NissinenVolovik2019}.  The invariant is equal to the second Chern number, and therefore reduces to an integral over the physical BZ. Time-reversal invariance and electric charge conservation suffices for non-trivial $N_{\theta}$ but are not necessary. With some other protecting symmetry $K$ it reads 
\begin{equation*}
    \begin{split}
N'_{\theta} = \frac{1}{24\pi} \int_{\omega=0}& d^{3}{\bf p} {\rm Tr}[\epsilon^{ijk} K (G^{-1} \partial_{p_i} G)\\ &\times (G^{-1} \partial_{p_j} G)(G^{-1} \partial_{p_k} G)],
\end{split}
\end{equation*}
where $K$ is the operator representation of the symmetry transformation. Examples of such topological invariant are provided by the superfluid $^3$He-B and Standard Model of particle physics \cite{Volovik2010}, when they are considered on the lattice. In other words, whereas in Eq.~\eqref{Nw1} ${\bf p}$ is fixed and the integral goes over the frequency, for the dual invariant the frequency is fixed and the integral goes over the momenta. 

Combined with the protecting symmetry, the theta term mod $\pi$ implies the protected boundary modes on the surfaces of the insulator \cite{Qi2008}. The invariant $N_{\theta}(x,t)$ is not quantized in general, however, and can be non-integer mod $\pi$ for solitonic configurations \cite{Vayrynen2011, Mulligan2013}, see also \cite{Kurkov2020}. 

Next we discuss non-trivial topological crystalline responses that are tantamount to extra crystalline topological conservation laws, featuring the elasticity tetrads, in addition to electric charge conservation (gauge invariance). These also imply protected boundary modes {in associated crystal directions}.}

\subsection{Anomalous QHE in 3D topological insulators}

In particular, the elasticity tetrads are important {in the field theory description} of the intrinsic (without external magnetic field) quantum Hall effect in 3D topological and axion insulators \cite{qhenote2}. The corresponding topological {response contains the elasticity tetrad as a dynamical lattice gauge field combined to the electromagnetic gauge field with the Chern-Simons topological term} \cite{Halperin1987, Matsuyama1987, Ishikawa1986, Kaplan1993,NissinenVolovik2019}:
\begin{eqnarray}
S[A,A,E]={\frac{1}{8\pi^2}} \sum_{a=1}^3N_a  
\int d^4 x~ E^{~a}_{\mu} \epsilon^{\mu\nu\alpha\beta} A_\nu \partial_\alpha A_\beta\,.
\label{action}
\end{eqnarray}
 {The response resulting from this action is {topologically} non-trivial and} the prefactor is expressed in terms of the topological {charges in momentum space}. {This implies chiral fermion modes on the boundary, relevant for the  3D QHE along $N_a\mathbf{E}^a$ (see Eq.~\eqref{ConductivityVariation} below), as well as on dislocations \cite{Halperin1987}. For superconductors $\tilde{A}_\mu = A_\mu-\partial_\mu\phi$, where $\phi$ is the symmetry breaking phase mode, can enter \cite{Stone2004} leading to chiral Majorana modes instead.} The three independent integer quantized coefficients $N_a$ are expressed in terms of integrals of the Green's functions in the energy-momentum space \cite{Matsuyama1987, Ishikawa1986}:
\begin{eqnarray}
N_a=\frac{1}{8\pi^2}\epsilon_{ijk} \int_{-\infty}^{\infty} d\omega\int_{\rm {BZ}} dS_a^i
\nonumber
\\
{\rm Tr} [(G\partial_{\omega} G^{-1}) (G\partial_{p_j} G^{-1}) ( G\partial_{p_k} G^{-1})]\,.
\label{Invariants}
\end{eqnarray}
Here the momentum integral is over the 2D torus --- the 2D  boundary $\bf{S}_a$ of the elementary cell of the 3D  reciprocal lattice, see Fig. \ref{fig:invariants}.

For a simple, {say}, {orthorhombic lattice in Fig. \ref{fig:invariants}}, the topological charge describing the QHE in, say, the $(x,y)$-plane  is $N_z$. It is the integral in the 
$(p_x,p_y)$ plane of the elementary cell of the reciprocal lattice at fixed $p_z$:
\begin{eqnarray}
N_z(p_z)=\frac{1}{4\pi^2} \int_{-\infty}^{\infty} d\omega\int_{{\rm BZ}} dp^x dp^y
\nonumber
\\
{\rm Tr} [(G\partial_{\omega} G^{-1}) (G\partial_{p_x} G^{-1}) ( G\partial_{p_y} G^{-1})]\,.
\label{Invariant_z}
\end{eqnarray}
{This integral in gapped crystalline insulators with AQHE does not depend on $p_z$, signaling the {quantized} response \eqref{action}}.

While in 2D crystals the topological invariant describes the quantization of the Hall conductance,  the topological invariants $N_a$ in 3D crystals describe the quantized  response of the Hall conductivity to deformation \cite{Halperin1987}:
\begin{equation}
\frac{d\sigma_{ij}}{dE^{\ a}_k} =\frac{e^2}{2\pi h}\epsilon_{ijk}  N_a\,.
\label{ConductivityVariation}
\end{equation}
The presence of the reciprocal lattice vector $E^{\ a}_k$ of dimension $1/[l]$ leads to the {correct dimensions of the 3D conductivity, as expected}.

\subsection{Polarization and flat bands in 3D topological insulators}

{The three topological invariants $N_a$ responsible for the 3D QHE are expressed in terms of integrals over three {planar} cross sections of the elementary cell of the three-dimensional reciprocal lattice, specified by perpendicular lattice directions. In three dimensions, there is another class of topological invariants represented by three invariants $N^a$ in terms of line-integrals along vectors of the reciprocal Bravais lattice {that couple to perpendicular planes as in Fig.~\ref{fig:invariants}}}. Such a line forms a closed loop {in the crystal that can, for example,} accumulate a Zak phase $\pi$, see e.g. Refs.~\onlinecite{Heikkila2011a,Heikkila2011} and \onlinecite{Rhim2017, Murakami2020}. 

The invariants $N^a$ 
are {related to a} topological {response} that can be considered dual to the action (\ref{action}), where one gauge field  $A_\mu$ is substituted by the tetrad gauge field. This is given by the following topological term in the action
\begin{equation}
S[A,E,E]= \sum_{a=1}^3 {\frac{N^a}{8\pi^2}} \epsilon_{abc} \int d^4 x E^b_\mu E^c_\nu  \epsilon^{\mu\nu\alpha\beta} \partial_\alpha A_\beta\,.
\label{actionPolarization}
\end{equation}
Since the term (\ref{actionPolarization}) is linear in the electric field ${\bf{\cal E}}=\partial_t{\bf A} - \nabla A_0$, 
 three invariants $N^a$ ($a=1,2,3$) characterize the topological polarization $\delta S[A,E,E]/\delta {\bf{\cal E}}$ along three directions. 
It leads to the induced boundary charges from the bulk, {in addition to modes bound on dislocations \cite{Vishwanath2019, NissinenVolovik2019}}. They are described by the action, assuming constant $A_{\mu}$ along $a$ at the boundary for simplicity, 
\begin{align}
S_{\rm bndry}[A,E,E] = \frac{\Delta N^a}{8\pi^2} \epsilon_{abc} \int_{\rm bndry} d^{3} x  E^b_\mu E^c_\nu  \epsilon^{\mu\nu\alpha}  A_\alpha. \label{eq:bndry}
\end{align}
It describes the {surface} polarization charge density coupling to $A_0$, and $\Delta N^a$ is the (integrated) bulk-boundary jump in $N^a$ and the integral is perpendicular to the direction $a$. Similar to the case of Eq.~\eqref{eq:volume_charge}, the static elasticity tetrads single out only the $A_0$ term. {For superconductors, the combination $A_0-\partial_t \phi$ can enter \cite{Stone2004}, with Majorana modes from polarization \cite{BudichArdonne2013}. Moreover, the boundary theory can be anomalous when considered without the associated bulk \cite{Oshikawa2000, Vishwanath2019}}.
 
{From the comparison of the polarization to the Zak phase, see e.g. {Refs. \cite{BudichArdonne2013, Rhim2017, Murakami2020} for insulators, superconductors and \cite{Heikkila2011a,Heikkila2011}} for gapless systems}, we conclude that in some cases the invariants $N^a$ can be written {simply} in terms of an effective Hamiltonian $H({\bf p})= 1/G({\bf p},\omega=0)$, which is the inverse of the Green's function at zero frequency.  {The polarization invariant can be more generally linked to  the semi-classical expansion for the momentum space invariants discussed in Ref. \onlinecite{NissinenVolovik2019}}. Here we assume that the insulator is $PT$ symmetric, i.e. obeys the combination of time reversal and space inversion symmetries, and thus the $PT$ operation commutes with the Hamiltonian. It is important that the operator  $PT$ is local in momentum space (see also \cite{Schnyder2016, Schnyder2018}), so that we can write the invariant in terms of an effective Hamiltonian. In particular, for an orthorhombic lattice the invariant is
\begin{align}
N^z(p_x,p_y) 
&=\frac{1}{2\pi i} {\rm Tr} \, \left[PT \oint dp_z   H^{-1} \partial_{p_z} H\right]\,.
\label{invariant}
\end{align}
Similar  to the invariant $N_z(p_z)$, which does not depend on $p_z$ in insulators, the invariant $N^z(p_x,p_y)$ does not depend on the transverse momenta ${\bf p}_\perp$ in the gapped systems (insulators or superconductors).

In non-interacting $PT$-symmetric insulators\cite{Murakami2020} and superconductors\cite{BudichArdonne2013} these invariants  determine the Berry phase change along the loop (the Zak phase), which is $2\pi N^a$. In nodal line semimetals the non-zero Zak phase  produces  zero-energy  surface states, which form a flat band \cite{Heikkila2011a,Heikkila2011}. In {gapped crystalline} insulators, where the invariants do not depend on $p_\perp$, the flat band occupies the whole Brillouin zone on the corresponding boundaries of the sample.  Note that the exact flatness of these surface bands rely on a chiral symmetry often present especially in nodal line superconductors \cite{Schnyder2011}, but also in approximative descriptions of nodal line semimetals. \cite{Heikkila2011a} This symmetry is not necessary for the stability of the nodal lines \cite{Burkov2011,Heikkila2018}, but in its absence the surface states become "drumhead" states with some dispersion. {The same is expected for the flat bands of the crystalline insulator \cite{Murakami2020}.}

\subsection{Dual invariants and quantized electric polarization response}

{To gain insight to the dual responses, including the polarization,} let us consider for simplicity an orthorhombic crystal with an electric field along $z$. {From Fig. \ref{fig:invariants},} the invariants $N^a$ can be considered as {geometric} duals to the three invariants $N_a$ in the crystalline lattice. While the invariant $N_3$ is an integral over the surface formed by two vectors, ${\bf E}^1\wedge {\bf E}^2$, the invariant $N^3$ is an integral on the path along the vector ${\bf E}^3$. They respectively couple to the tetrads $\mathbf{E}^3$ and $\mathbf{E^1} \wedge \mathbf{E}^2$ in the response.

{We now focus explicitly on the polarization}. Then the appropriate part of the action contains the invariant $N^3$:
\begin{equation}
S[A,E,E]=\frac{N^3 }{4\pi^2}\int d^4 x ({\bf E}^1 \times  {\bf E}^2) \cdot   {\bf{\cal E}}= \frac{N^3}{4\pi^2}\int d^4 x \,{\bf S}^{12} \cdot   {\bf{\cal E}} \,,
\label{actionPolarizationRect}
\end{equation}
where ${\bf S}^{12} $ is the area of the 2D BZ in the plane perpendicular to the normal of the considered boundary. 

Electric polarization is determined as the response of the action to the electric field ${\bf{\cal E}}$  in the limit of infinitesimal electric field, ${\bf{\cal E}}\rightarrow 0$. From Eq.~(\ref{actionPolarizationRect}) it looks that for the topological insulator with $N^a\neq 0$, the polarization is non-zero  in zero electric field, which is however forbidden by parity symmetry, or by the $PT$ invariance. In fact, it is forbidden for the infinite sample, while in the presence of boundaries this is possible, since boundaries violate parity symmetry, {similar to the time-reversal symmetry and surface modes with theta term}. In the presence of two boundaries there are two degenerate ground states with opposite polarization. In one state the positive electric charges are concentrated on the upper boundary (with electric charge $+|e|/2$ per one state in the flat band), and the negative charges are on the lower boundary. In the other degenerate state the polarization is opposite. The first state is obtained as a response to the electric field ${\cal E}_z \rightarrow +0$, while the second state is obtained in the limit ${\cal E}_z \rightarrow -0$. This means that the integer topological polarization can be considered as the difference in polarization, when the electric field changes sign.

Recent calculations of the topological polarization in nodal loop semimetals have been done in Ref.~\onlinecite{Nomura2018}.
We consider this for {crystalline} topological insulators where {the response is quantized in terms of the elasticity tetrads}.  Similar to the response of the QHE to deformations in Eq.~(\ref{ConductivityVariation}), which is quantized in {crystalline} topological insulators in terms of invariants $N_a$, the response of the topological polarization to {strain} is quantized in terms of the invariants $N^a$.
From Eq.~(\ref{actionPolarizationRect}) it follows that {the {quantized} response corresponding to the polarization $P^i = \delta S /\delta \mathcal{E}_i\vert_{\mathcal{E}=0}$} is the deformation of the cross sectional area in the reciprocal lattice: 
\begin{equation}
 \frac{dP^i}{dS^{ab,k}} =  \frac{1}{4\pi^2}\delta^i_k \epsilon_{abc}N^c\,.
\label{PolarizationResponse}
\end{equation}
For the simple orthorhombic crystal and for polarization along $z$ {this becomes}
\begin{equation}
{\frac{dP^z}{dS^{12}}}=\frac{1}{4\pi^2} N^3\,.
\label{PolarizationResponseRect}
\end{equation}
The {quantized variation} of the polarization with respect to deformation is an example of a well defined  "differential" polarization \cite{Resta1994,Resta2010}. Note that the polarization itself is not quantized, depending on {(the surface spanned by)} the reciprocal lattice vectors, but its derivative with respect to deformation in Eq.~(\ref{PolarizationResponse}) is quantized. 

\section{Polarization and flat band in a numerical model}

In 3D topological insulators, the same invariant $N^a$ hence {implies} both the flat band on the surface of the material 
and the topological polarization {in the bulk response}. {In general terms, this is an example of bulk-boundary correspondence or} anomaly inflow from the bulk to the boundary, {as discussed above}.

More concretely, this follows since each ${\bf p}_\perp$ the system represents a {1+1d} topological insulator, and thus for each ${\bf p}_\perp$ there should be a zero energy state on the boundary.  Thus for the topological insulators with nonzero $N^c$ the flat band exists on the surface for all ${\bf p}_\perp$. This is distinct from nodal line semimetals, where the region of the surface flat band is bounded by the projection of the nodal line to the boundary. The topological insulator phase can be obtained when the Dirac loop is moved to the boundary of the BZ.

\begin{figure}[h]
\centering
\includegraphics[width=\columnwidth]{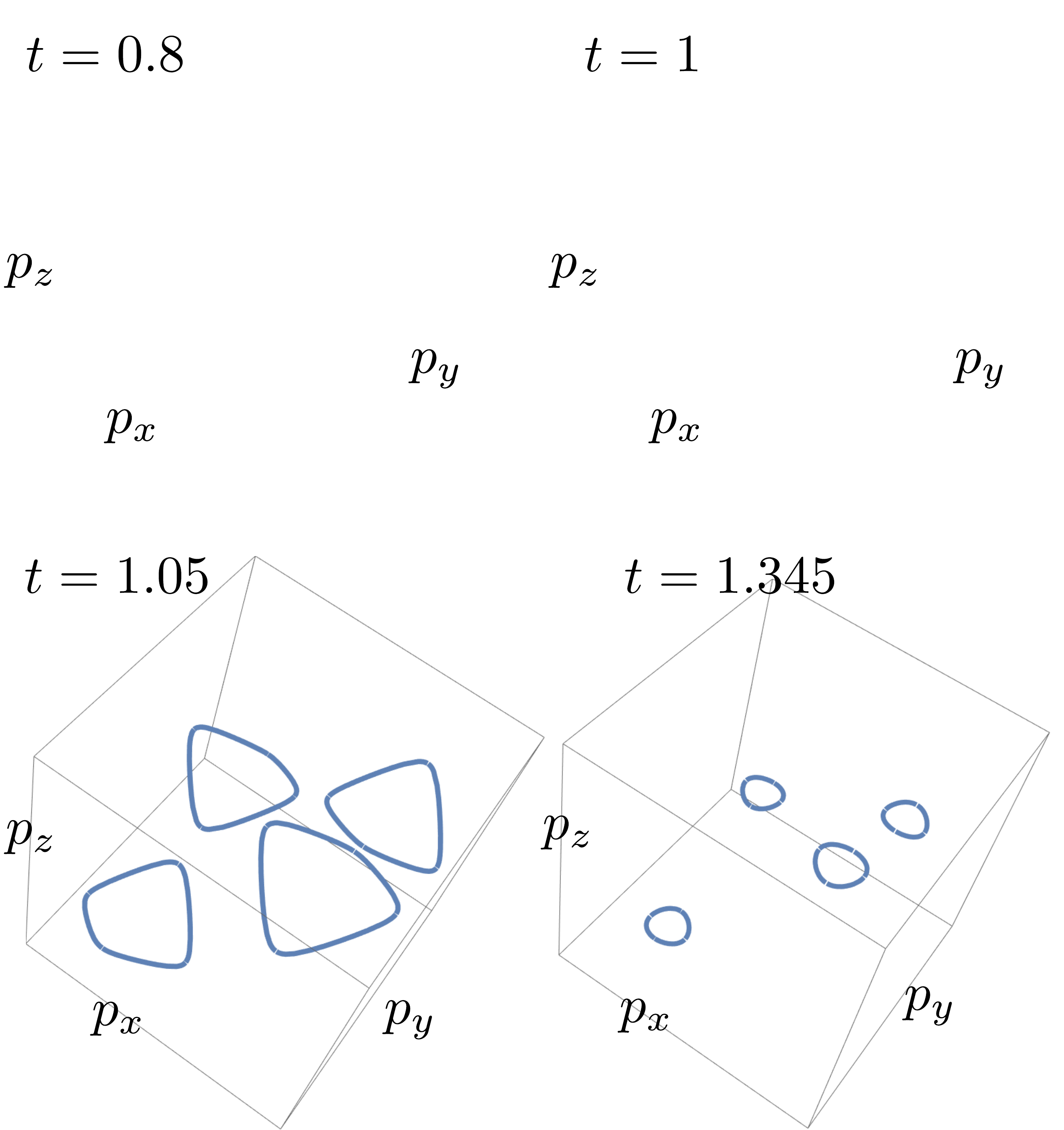}
\caption{Nodal lines for different parameters $t$. In top figures showing the spiral lines up to $t=1$, the first Brillouin zone is shown as a box. In bottom figures, only the first Brillouin zone is plotted. }
\label{fig:nodallines}
\end{figure}

\begin{figure}[h]
\centering
\includegraphics[width=\columnwidth]{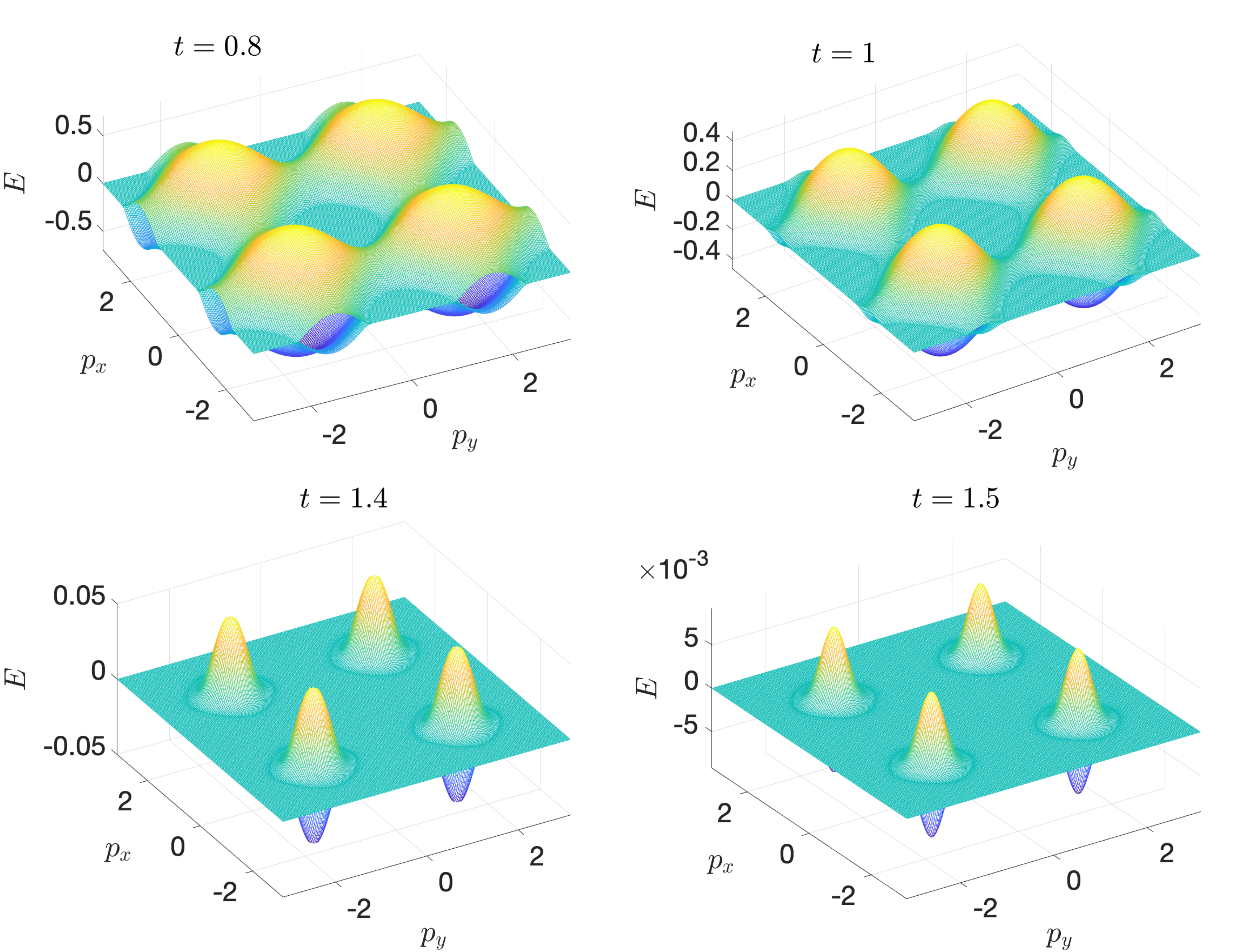}
\caption{Two lowest-energy eigenstates near $E=0$, showing how for $t<\sqrt{2}$ the flat bands extend through part of the first Brillouin zone, and for $t>\sqrt{2}$ across the entire B.Z. In the figure with $t=1.5$, the shown finite energy is associated to the finite number of layers in the simulation (note the energy scale that is lower than in other plots). All plots are computed with $N=51$ layers.}
\label{fig:flatbands}
\end{figure}

This can be {verified} using an extension of the model $H = \left(\begin{matrix} & f \\ f^* & \end{matrix}\right)$ in Ref.~\onlinecite{Heikkila2011a} with $f=\sin p_x + i \sin p_y - te^{-i p_z}$, i.e.
the Hamiltonian in the limit of infinite number of layers is
\begin{equation}
H = \sigma_x (\sin p_x - t \cos p_z) +\sigma_y(\sin p_y - t \sin p_z)\,.
\label{Hamiltonian}
\end{equation}
For low enough $t$, the nodal line can be found at the momenta $p_x,p_y,p_z$ that simultaneously nullify the coefficients of $\sigma_{x,y}$. This model has three different phases depending on the value of the coefficient $t$ as illustrated in Figs.~\ref{fig:nodallines} and \ref{fig:flatbands}. For $t<1$, the first Brillouin zone contains four spiral lines, one inside it, others going through the Brillouin zone boundaries. In this case there are surface flat bands at the projection of the spirals to the surfaces. At $t=1$ these lines touch and cut each other to form closed nodal line loops when $1<t<\sqrt{2}$. The projection of these loops to the surface still mark the boundaries of the surface flat bands. Finally, for $t=\sqrt{2}$ the loops shrink into four nodal points and vanish for $t>\sqrt{2}$ in which case the system forms a topological insulator. In this case the flat band extends throughout the 2D Brillouin zone of the transverse momenta. This behavior is qualitatively similar to that found for the slightly more complicated model of rhombohedrally stacked honeycomb lattice.\cite{Heikkila2018}
 
 The topological invariant \eqref{invariant} is in Eq.~5 of Ref.~\onlinecite{Heikkila2011a}, where the $PT$ operator is played by $\sigma_z$. In terms {of the unit vector of Pauli matrices in the Green's function $\hat{\mathbf{g}}(\mathbf{p},\omega)\equiv \frac{\mathbf{G}}{\vert \mathbf{G} \vert}$} the invariant is in Eq.~(8) of Ref.~\onlinecite{Heikkila2011a}:
\begin{equation}
N^3({\bf p}_\perp)=\frac{1}{4\pi}\oint_{-\pi/a}^{\pi/a} dp_z
  \int_{-\infty}^{\infty} d\omega \hat{\bf g}\cdot \left( \frac{\partial \hat{\bf g}} {\partial p_z} \times \frac{\partial \hat{\bf g}} {\partial \omega} \right)\,.
\label{invariant2}
\end{equation}
In the case of the Hamiltonian in Eq.~\eqref{Hamiltonian}, in the nodal line phase corresponding to $t<\sqrt{2}$, $N^3({\bf p}_\perp)$ is non-zero inside the projection of the nodal lines to the 2D space ${\bf p}_\perp$ and zero outside it. On the other hand, in the topological insulator phase with $t>\sqrt{2}$, $N^3({\bf p})=1$ for all transverse momenta.

For a finite number of layers the Hamiltonian matrix is
\begin{equation}
H_{ij}=(\sigma_x \sin p_x  +\sigma_y\sin p_y)\delta_{ij} - t(\sigma^+ \delta_{i,j+1} +\sigma^- \delta_{i,j-1})\,.
\label{HamiltonianDiscrete}
\end{equation}
This can be used to compute the spectrum shown in Fig.~\ref{fig:flatbands}. Moreover, using the (spinor) eigenstates $\phi_n(j,p_x,p_y)$ of the finite-system Hamiltonian corresponding to eigenenergy $\epsilon_n$, we also get the charge density at layer $j$
\begin{equation}
    \rho_j = \rho_0-e \sum_n \int_{BZ} \frac{d^{(2)} p}{(2\pi)^2} f(\epsilon_n) \phi_n(j,p_x,p_y)^\dagger \phi_n(j,p_x,p_y).
    \label{eq:chargedensity}
\end{equation}
Here the integral goes over the 2D Brillouin zone of size $S^{12}$ of the transverse momenta, $f(\epsilon)$ is the Fermi {distribution} and $\rho_0=e S^{12}/(4\pi^2)$ ensures a charge neutral situation at zero chemical potential. We calculate everything at at zero temperature.

In a given electric field, the polarization can be computed as
\begin{equation}
    P^z = \frac{1}{2}\sum_{j=1}^{N} \rho_j \mathrm{ sgn}(j-N/2),
    \label{eq:polarization}
\end{equation}
We calculate this polarization in the case of an applied electric field similarly as in Ref.~\onlinecite{Nomura2018} (see Appendix for details) \cite{numericsnote}. We mostly concentrate on the case of negligible screening, i.e., disregard the back-action of the charge density to the electric field. This corresponds to the limit $\alpha \rightarrow 0$ in Ref.~\onlinecite{Nomura2018}. The results are shown in Fig.~\ref{fig:polarization}. Due to the presence of the flat bands, a small electric field leads to a charge density that is antisymmetric with respect to the center of the system (the average charge hence vanishes), i.e., a non-zero charge polarization. This polarization jumps rather abruptly as a function of the sign of the electric field. The size of the jump is integer
\begin{equation}
    P^z(\mathcal{E}_z>0)-P^z(\mathcal{E}_z<0) =  e \frac{\Omega_{\rm FB}}{4\pi^2},
    \label{polarizationjump}
\end{equation}
where $\Omega_{\rm FB}$ is the area of the flat band in momentum space. In the topological insulator phase $t>\sqrt{2}$ the size of the flat band becomes equal to the size of the 2D Brillouin zone, $\Omega_{\rm FB} = S^{12}$, and hence we get the result of Eq.~\eqref{PolarizationResponseRect}.

\begin{figure}[h]
\centering
\includegraphics[width=\columnwidth]{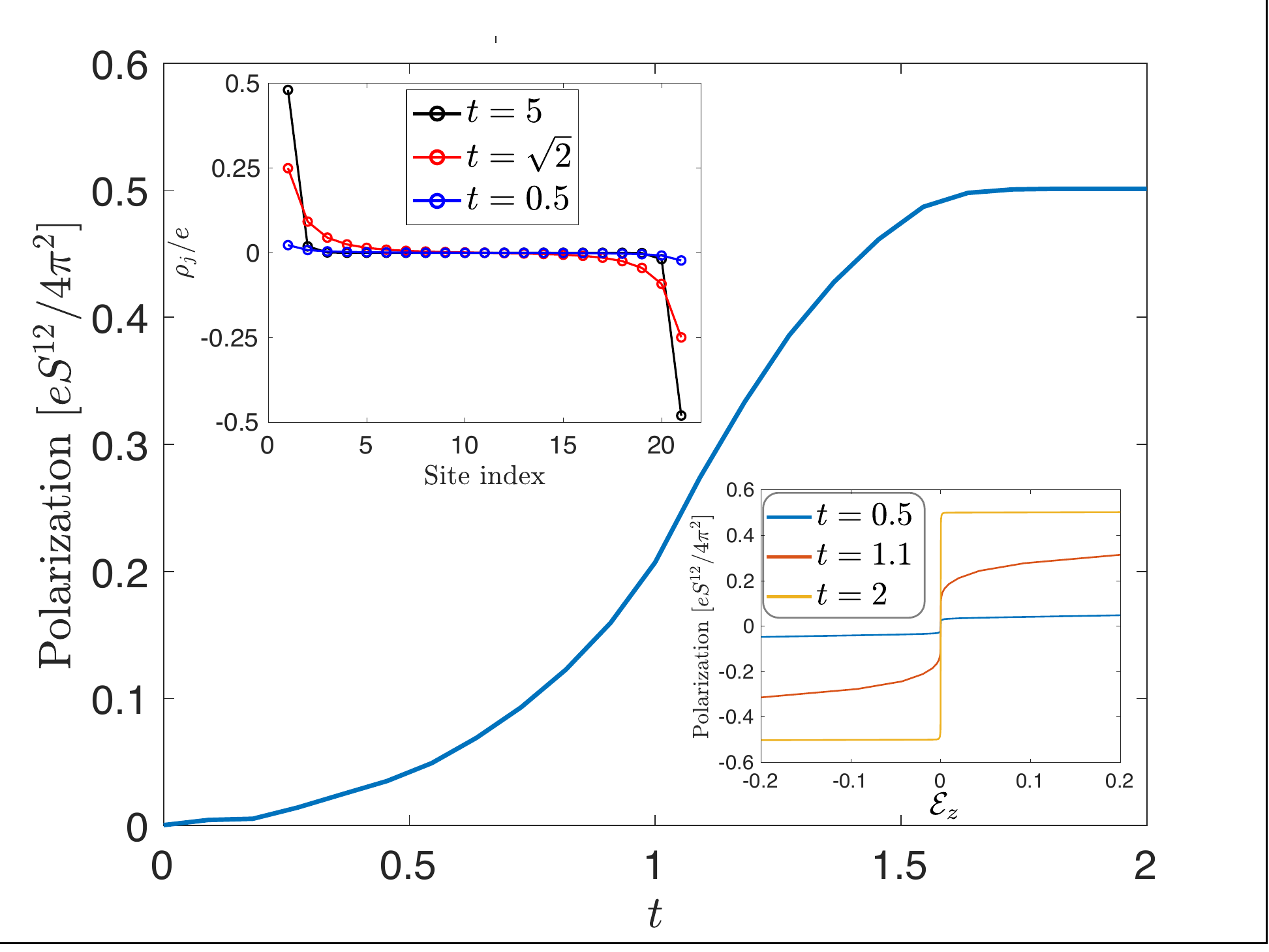}
\caption{Polarization as a function of the hopping parameter $t$ driving two Lifshitz transitions from two types of nodal line semimetals at $t<1$ and $1<t<\sqrt{2}$ to a topological insulator phase at $t>\sqrt{2}$. Upper inset: charge density for a few parameter values indicated in the legend. Lower inset: polarization as a function of the electric field $\mathcal{E}_z$. If not specified otherwise, the figures are calculated with $N=21$ layers, and with electric field $\mathcal{E}_z=0.1$.}
\label{fig:polarization}
\end{figure}

Note that the model described here contains a chiral symmetry: $H$ anticommutes with the $PT$ symmetry operator $\sigma_z$. Such chiral symmetries are typically not encountered in crystal lattices, but they may be approximate symmetries in their model Hamiltonians (for the case of rhombohedral graphite, see Ref.~\onlinecite{Kopnin2013}). Chiral symmetry breaking terms do not destroy the surface states, but in their presence the surface states become drumhead states with a non-zero bandwidth $\delta \epsilon$. In this case the polarization no longer contains an abrupt jump as a function of the field, but the jump has a finite width. Nevertheless, the size of the jump remains the same as in Eq.~\eqref{polarizationjump}. 

\section{Conclusion and Outlook}

We discuss the topological responses in three-dimensional crystalline insulators (and superconductors) in terms of dual pairs of invariants and the elasticity tetrads. We focus on the topological polarization response and the associated flat bands at the boundaries. This polarization response and modes are distinct from e.g. the bulk theta term in time reversal invariant topological insulators, in that they are protected only by the associated (weak) crystalline symmetries. We discuss the relation of polarization to the other possible invariants in three dimensions and show the explicit momentum space invariant linking the bulk and boundary. We also demonstrate the formation of the topological polarization in the case of an example Hamiltonian specified in Eqs.~(\ref{Hamiltonian},\ref{HamiltonianDiscrete}). 

{In more detail}, the one-dimensional topological invariant (the Zak phase) in Eqs. (\ref{invariant}) and (\ref{invariant2}) describes two {related} phenomena: the topological response of polarization to the strain and the {surface} flat band.
This demonstrates that the {bulk} topological polarization {implies} the filling of the
zero energy surface states {{and} vice versa, constituting an example of bulk-boundary correspondence and the associated anomaly inflow}. Notably, the bulk polarization response is a total derivative. {Using a simple model}, we {explicitly verified} that the response of the polarization to the properly defined deformations is quantized, see Eq.~(\ref{PolarizationResponse}), and {that the corresponding surface flat band is present throughout the whole BZ}. This is distinct from the nodal line semimetals, where there is also a flat band, but where this flat band occupies only part of the surface BZ. As a result there is no {bulk} quantization. 
The {surface polarization} and theory become anomalous in terms of a mere two-dimensional description, and they have to be discussed in the context of the bulk-boundary correspondence, including the gapless fermions. {However}, the polarization difference and derivative with respect to the deformation becomes quantized {precisely} when {the nodal loop moves to the boundaries of the BZ and annihilates}, forming a gapped topological insulator. 

This situation is very similar to that in the {AQHE}, {implying the presence of protected chiral edge modes}. In 3D topological {crystalline} insulators it is the derivative 
of the Hall conductivity which is quantized \cite{NissinenVolovik2019} {and well-defined}.  In the Weyl semimetals such quantization is absent, {implying the chiral anomaly from the gapless fermions}, but is restored when the Weyl nodes move to the boundaries of the BZ and annihilate forming a topological insulator, {see e.g. \cite{Zyuzin2012, Hughes2016, NissinenVolovik2019}}.

Systems with flat bands are strongly susceptible to interaction induced broken symmetry phases such as superconductivity \cite{Kopnin2011}. There,  the (mean field) transition temperature $T_c$ is proportional to the volume of the flat band, if the flat band is formed in the bulk \cite{Khodel1990}, or to the area of the flat band if it is formed on the surface of the sample \cite{Heikkila2011}. Topological insulators have a larger area of the flat band compared with the flat bands on the surface of nodal line semimetals, and thus they may have a higher $T_c$. {This is in contrast for example to the Moir\'e superlattice in magic angle twisted bilayer graphene, where the flat band extends across the first Brillouin zone of the superlattice \cite{SuarezMorell2010,Bistritzer2011,Peltonen2018}. At the magic angle the superlattice unit cell contains a large number $N \sim 10
^4$ of atoms, implying a rather small flat band with area $\sim 1/(Na^2)$, where $a$ is the graphene lattice constant. Nevertheless, the recent measurements \cite{Cao2018,Yankowitz2019} indicate superconductivity with a $T_c$ around a few K.} This means that topological insulators with much larger flat bands may be included in the competition whose final goal is room-temperature superconductivity. Other current participants in the race are hydrogen-rich materials, such as H$_3$S,  LaH$_{10}$, Li$_2$MgH$_{16}$, YH$_6$, etc. \cite{Heil2019, Oganov2019, Kong2019, Drozdov2019, Grockowiak2020} In these systems, the large transition temperature results from the large vibrations of light hydrogens, which increase the electron-phonon coupling. The contributions of flat band and vibrations would ideally be combined. The manipulation and control of acoustic vibrations in insulators (which represent massive and masless "gravitons" in terms of the lattice metric and elasticity tetrads with elastic energy, respectively \cite{Wetterich2012}) is not an easy task. But if the surface flat band of the insulator is in contact with hydrogen-rich material, then the electron-phonon interaction between phonons in hydrides and electrons in flat band may conspire in increasing $T_c$ even further. 

Lastly, we note that a phase with periodic string-like order parameter in spin
chains was recently found to lead to topological flat bands of Majoranas
\cite{Pandey2020}. This is probably related to the polarization in the
crystalline superconductors. On the other hand, another recent flat-band work describing lattices of fermions with random
interactions \cite{Patel2019,Volovik2019} is rather related to the Khodel-Shaginyan Fermi
condensate \cite{Khodel1990}.

This work has been supported by the European Research Council (ERC) under the European Union's Horizon 2020 research and innovation programme (Grant Agreement No. 694248) and the Academy of Finland (project No. 317118).

\appendix

\section{Details of the numerics}

Figure \ref{fig:nodallines} is produced by a parametric plot exhibiting the simultaneous solutions to the equations
$$
\sin(p_x)=t \sin(p_z); \sin(p_y)= t\cos(p_z).
$$
Figure \ref{fig:flatbands} is obtained by constructing the $2N \times 2N$ matrix corresponding to Eq.~\eqref{HamiltonianDiscrete}. The plotted quantity corresponds to the two center eigenvalues, which are the lowest-energy eigenstates at $\mu=0$ for the particle-hole symmetric Hamiltonian. 

Figure \ref{fig:polarization} finds the eigenstates of the Hamiltonian $H_{ij}-\mu_j \delta_{ij}$ with a layer dependent potential $\mu_j$. To mimic an electric field in the direction perpendicular to the layers, we follow Ref.~\onlinecite{Nomura2018} and choose 
$$
\mu_j = {\mathcal E}_z (j-N/2).
$$
Using the resulting eigenstates and -energies, we then calculate the charge density Eq.~\eqref{eq:chargedensity} and polarization Eq.~\eqref{eq:polarization}. Note that this approach neglects the changes into $\mu_j$ that would come from solving the Poisson equation. It hence corresponds to the limit $\kappa \rightarrow \infty$ or $\alpha \rightarrow 0$ in Ref.~\onlinecite{Nomura2018}. The case of a finite $\kappa$ would lead to a possibility of broadening of the polarization step, but would not affect the size of the step. Moreover, we have studied the effects of chiral symmetry breaking terms (that do not anticommute with $\sigma_z$). They lead to a non-vanishing bandwidth of the surface states similar to what happens in rhombohedral graphite \cite{Kopnin2013}. As long as such terms are weak, they only broaden the polarization jump but do not change its overall magnitude.

\bibliography{refs}

\begin{thebibliography}{67}
\bibitem [{qhe({\natexlab{a}})}]{qhenote1}%
  \BibitemOpen
  \href@noop {} {} ({\natexlab{a}}),\ \bibinfo {note} {the discussion in this
  manuscript is for spinless fermions. In other words, we do not concentrate on
  the distinction between time-reversal symmetry breaking quantum Hall effect
  and the spin quantum Hall effect relevant in the presence of time reversal
  symmetry.}\BibitemShut {Stop}%
\bibitem [{qhe({\natexlab{b}})}]{qhenote2}%
  \BibitemOpen
  \href@noop {} {} ({\natexlab{b}}),\ \bibinfo {note} {the discussion in this
  manuscript is for spinless fermions. In other words, we do not concentrate on
  the distinction between time-reversal symmetry breaking quantum Hall effect
  and the spin quantum Hall effect relevant in the presence of time reversal
  symmetry.}\BibitemShut {Stop}%
\bibitem [{num()}]{numericsnote}%
  \BibitemOpen
  \href@noop {} {}\bibinfo {note} {The codes needed to reproduce
  Figs.~\ref{fig:flatbands} and \ref{fig:polarization} will be linked to a
  published work.}\BibitemShut {Stop}%
\end{thebibliography}

 \end{document}